# The Synthesis and Electrical Transport of Ligand-Protected Au$_{13}$ Clusters

Zhongxia Wei,[1] Wanrun Jiang,[2] Zhanbin Bai,[1] Zhen Lian,[1] Zhigang Wang,[2] and Fengqi Song[1, a)]

[1]*National Laboratory of Solid State Microstructures, Collaborative Innovation Center of Advanced Microstructures, and College of Physics, Nanjing University, Nanjing 210093, P. R. China*

[2]*Institute of Atomic and Molecular Physics and Jilin Provincial Key Laboratory of Applied Atomic and Molecular Spectroscopy, Jilin University, Changchun 130012, China.*

---

a) Corresponding author. Prof. Dr. Fengqi Song. Email: songfengqi@nju.edu.cn

Tel. +86-25-83592701. Fax. +86-25-83595535

Abstract: The ligand-protected $Au_{13}$ clusters have been synthesized by using meso-2,3-imercaptosuccinic acid as the reducing and stabilizing agent. Transmission electron microscopic analysis shows a size distribution of 1.4 ± 0.6 nm. Optical spectrum shows an absorbance peak at 390 nm. The electrical transport measurement devices are fabricated using the electro-migration method. Coulomb blockade is observed at the temperature of 1.6 K, revealing the formation of the tunneling junction. The Coulomb oscillation's on/off ratio is nearly 5. Three peaks are extracted in the dI/dV data and attributed to the energy levels of $Au_{13}$ clusters, gapped by about 60 meV. First principle calculations are carried out to interpret the energy diagram.

## I. INTRODUCTION

Atomically precise ligand-protected clusters are the aggregates with defined number of atoms, typically Au or Ag, surrounded by a protective layer of ligands, with number-dependent atomic structures and exhibiting many special optical, electrical, catalytic and magnetic properties[1-4]. In some applications towards quantum memory, metallic atomic clusters have proven advantageous over the organic molecules due to their greater environmental stability, higher density of states and mature fabrication techniques[5-7]. Comparing to other metal clusters, tiny gold nanoclusters, which fulfill the first geometrical shell closing: 13, 19, 43, 55, etc. are expected to be important building blocks of cluster devices due to their novel chemical, electronic and optical properties[8,9]. $Au_{13}$ clusters are the smallest magic number clusters and have attracted much attention. Plenty of work has been done on the $Au_{13}$ clusters regarding of its synthesis, structure, electronic structures, etc.[10-19] However, the fabrication of $Au_{13}$-based cluster devices and their electrical transport have not been reported yet.

In this work, we have fabricated the cluster devices based on ligand-protected $Au_{13}$ clusters. Gold clusters $Au_{13}$ are produced by the wet chemical method and the measurement devices are made by the electro-migration technique. Coulomb blockade phenomena and quantum level spacing were observed at 1.6K. To the best of our knowledge, this is the first electrical transport measurement on the $Au_{13}$ clusters.

## II. EXPERIMENTS AND RESULTS

The synthesis of the gold nanoparticles follows the route of Negishi *et al.*[19] Fig.1a illustrates the synthesis process. For this synthesis, 41.2mg (0.1mmol) of HAuCl4.4H2O was dissolved in 10mL water to which 36.8mg (0.2mmol) meso-2,3-imercaptosuccinic acid (DMSA, HO2CCH(SH)CH(SH)CO2H) was added under vigorous stirring at ambient conditions. The color of the solution changed from clear yellow to brown at the very beginning and turned dark yellow after 10 minutes. Thiolate complexes and unreacted DMSA molecules were removed by centrifugal filtration. DMSA was acting as the reducing and stabilizing agent during the process. To the aqueous solution of the Au:DMSA clusters, 10ml toluene solution of 54.7mg (0.1mmol) tetraoctyl ammonium bromide (TOABr, BrN(n-C8H17)4) was added to form two immiscible layers. After vigorous stirring at room temperature for 30 minutes, the gold clusters were extracted to the toluene phase to form the Au:DMSA-TOA clusters. The synthesized gold clusters were characterized by UV/vis spectrum and transmission electron microscope (TEM). The TEM specimens were prepared by depositing one or two drops of the sample solutions onto ultrathin carbon-coated copper grids. TEM images were acquired with a FEI Tecnai F20 transmission electron microscope operated at 200 kV. Optical spectra of the Au:DMSA clusters was recorded by using a INESA 722N spectrometer.

Fig. 1b shows a typical TEM image of the synthesized Au:DMSA-TOA clusters. Most clusters under investigation display almost spherical shapes. According to Negishi *et al*.[19] $Au_{13}DMSA_8$ clusters are the most abundant products of the reaction. Fig.1c gives the size distribution of the gold clusters, which have a mean diameter of 1.4 nm with a deviation of 0.6 nm, agreeing with the previous reports on $Au_{13}$. The optical spectrum of the Au:DMSA clusters exhibits absorption onset at ca.500nm and peak structures ca. 390nm (Fig. 1d). This peak is in the similar position of those previous optical spectrum studies of ligand-protected $Au_{13}$ clusters[12,18].

The electrical transport properties of the synthesized gold clusters were investigated by incorporating the cluster into single-electron transistors (SETs). The SETs were fabricated by the electro-migration technique[20]. One or two drops of the organic gold cluster solution were deposited onto the gold electrodes and a gap of 1-2nm was produced by electro-migration. Gate tunable double barrier tunnel junction was formed when a single gold cluster was trapped in the gap and coupled to the source and drain electrodes. The sample was placed under vacuum at cryogenic temperatures. A subfemtoampere sourcemeter (Keithley 6430) was used to carry out the electrical transport measurement.

Fig.2a schematically illustrates the SET device structure. The electrodes are made by a standard e-beam lithography. A break-junction technique was then used to create a gap between these electrodes by the process of electro-migration[20]. The entire structure was defined on a $SiO_2$ insulating layer on top of a degenerately doped silicon wafer that serves as a gate electrode that modulates the electrostatic potential of the gold clusters. The devices can be modeled by the gold cluster controlled by three electrodes. Two electrodes (source and drain) are tunnel coupled to the cluster and electrical transport is allowed only between the cluster and these two electrodes.

Fig.2b presents the current-voltage (I-V) curves obtained at 1.6 K from the gold cluster at different gate voltages ($V_g$). As shown in the figure, each curve in Fig.2b shows a suppressed conductance near zero bias voltage followed by step-like current jumps at higher voltages. This suppression of conductance at low biases is a direct result of the charge addition energy which can be attributed to the Coulomb blockade, a signature of single electron transistors. As seen in Fig.2b, the I-V curves almost stay unchanged as the $V_g$ varied up to 11V. Note that the gate thickness is around 30nm, where such gate voltages have been high enough to tune the energy levels in previous single molecule devices[21-23]. Usually, the width of the Coulomb blockade can be changed by changing $V_g$. Such trivial $V_g$-dependence is attributed to the poor coupling between the gold cluster and the gate electrode, or the very high electronic density.

Further work will improve the devices' quality.

In Fig. 2c, we also plot the differential conductance dI/dV, as a function of bias voltage at 1.6 K for different gate voltages. Each dI/dV-V curve shows a peak near current steps in the corresponding I-V curves in Fig. 2b. Three differential conductance features can be seen at about V = -3, 20, 80 mV. Fine structures might be contained since there are some shoulder shapes. This is the contribution of discrete energy levels of gold clusters. All the peaks are the result of the resonant tunneling when one energy level of the gold cluster is aligned to the Fermi level of source or drain electrodes. The observed biggest energy gap of the two energy levels of the gold cluster can be derived from the spectrum to be about 60 meV. Besides, the Coulomb oscillation on/off ratio can be estimated from the spectrum to be nearly 5.

The transport features described above were not observed in devices when the gold cluster was not deposited on the electrodes. Besides, the devices exhibited similar conductance characteristics that are consistent with a single nanometer-sized object bridging two electrodes. Although the gold cluster could not be imaged directly in the device due to its small size, these experimental observations indicate that the gold cluster is responsible for the conductance features observed in the experiment.

Fig.2d shows the results of electrical current measurement through the gold cluster at 1.6, 6, 30, 100 and 150 K. The I-V curves observed at 1.6, 6 and 30 K show the nonlinearity at around zero bias voltage. Besides, the conductance increases as the temperature increase. At 100 and 150 K, the nonlinearity is smeared out, indicating no Coulomb blockade observed for the temperatures above 100 K. This is most probably because the devices' physical configuration has changed over the time of the different temperature measurements.

As the calculation model, DMSAs are used to passivates an $Au_{13}$ cluster in the constructed structure, in which 4 DMSAs lost their H atoms in both two mercapto groups and the other 4 keep one S-H bond per molecule, providing 12 S atoms with dangling bonds in total. This is in accordance to the experiments of Negishi *et al.*[19] where $Au_{13}(DMSA)_8$ is the most abundant product while mercapto groups in DMSAs would lose their H atoms in a proportion of 3/4 after the preparation procedure. Meanwhile, an $Au_{13}$ cluster with 12 outer-shell atoms and an $D_{4h}$ symmetry is chosen to be the initial core, because this configuration is reported to be an energy-preferred isomer in another eight-ligand protected condition[17]. The system presents synergistically rolled up ligands and the deformed $Au_{13}$ cluster after the rough relaxation. All outer-shell Au atoms are passivated by the Au-S bonds with DMSAs. Molecular planes of ligands lean towards the surface of the Au clusters. Six ligands formed an "equator" ring with an anticlockwise leaning orientation as shown by the view in Fig. 3a. Two ligands cover the corresponding bottom and top regions with opposite lean directions. The intermolecular hydrogen bonds between DMSAs may influence the ligand arrangement given the carboxyl groups in bottom and top ligands are close to those in equator ones.

The density of states (DOS) of this structure is calculated and shown in Fig. 3b. It is shown after the zero point shifted to the energy of the highest occupied molecular orbital (HOMO). Different from the reported DOS diagram of $Au_{13}$, this calculation is carried out in a very high resolution ~ 0.001eV, in order to compare with the experimental data with the source –drain voltage is up to around 0.1-0.5V normally. Four DOS peaks are shown in Fig. 3b, which correspond to states of the HOMO and three above unoccupied states respectively. As shown in the inset, the isosurfaces show the HOMO electrons are mainly distributed on two S atoms of one ligand with the morphology feature of the p atomic orbital. Hence the easier coupling of $Au_{13}$ with S-contained molecule is reasonable. The lowest unoccupied molecular orbital (LUMO) shows a similar distribution. LUMO+1 and LUMO+2 show the contributions both from S atoms and Au atoms, which may be attributed to the deformation of the Au cluster. The interval between HOMO and LUMO are narrower than that between LUMO an LUMO+1 while an even smaller interval can be found between the LUMO+1 and the LUMO+2. We note that the HOMO-LUMO gap is about 60 meV.

The experimental dI/dV curves (Fig.2c) can be compared with the DOS data (Fig. 3b), which can be interpreted as the energy levels of the cluster. Zero in Fig.2c is set to the Fermi level, while the zero assignment in the cluster energy levels is on debates, possibly due to the local electronic configuration and some inuniformity. One may see there is a small gap of around 0.06 eV in the DOS curves, which can be related to the blue and black peaks measured experimentally in Fig. 2c. We propose the diagram as shown in Fig. 3c to interpret the measurement in our SCDs. We consider the measurement of a $Au_{13}$ cluster with the energy gap of 60 meV (from calculation) as the initial stage of Fig. 3c, where the HOMO is positioned 3 meV below the Fermi level. The coupling strength of the left/right electrodes is set to 3:17, a reasonable value. While a negative bias voltage is applied, the HOMO contribute a dI/dV peak at -3meV. When the bias voltage is increased to a small positive value, the left electrode will touch the bottom level (HOMO), which contributes a dI/dV peak at 20 meV. When the bias voltage is higher, the LUMO will be touched with the contribution of the dI/dV at 67 meV. This leads to the experimentally observed three peaks.

III. CONCLUSIONS

In conclusion, ligand-protected $Au_{13}$ clusters were prepared by the reactions between $HAuCl_4$ and DMSA molecules. Measurement devices are made by the electro-migration method. The charge transport behavior of the gold clusters has been studied by the devices. Coulomb blockade is observed at the temperature of 1.6 K. The Coulomb oscillation's on/off ratio is nearly 5. The energy levels of the $Au_{13}$ cluster are revealed by the conductance peaks in the dI/dV spectrum. A possible structure and density of states of $Au_{13}(DMSA)_8$ cluster are calculated by DFT, in which the HOMO-LUMO levels contribute the experimentally measured differential conductance peaks.

ACKNOWLEDGEMENT

We would like to thank Prof. Sufei Shi for his help and we gratefully acknowledge the financial support of the National Key Projects for Basic Research of China (Grant Nos: 2013CB922103), the National Natural Science Foundation of China (Grant Nos: 91421109, 91622115, 11574133 and 11274003), the PAPD project, the Natural Science Foundation of Jiangsu Province (Grant BK20130054), and the Fundamental Research Funds for the Central Universities. We would also like to acknowledge the helpful assistance of the Nanofabrication and Characterization Center at the Physics College of Nanjing University. Z.W. also acknowledges the assistance of the High Performance Computing Center (HPCC) of Jilin University.

FIGURE CAPTIONS

**Figure 1. Synthesis and Characterization of Au$_{13}$ clusters**

 **(a)** Synthesis process of Au:DMSA clusters. **(b)** TEM image of the ligand-protected gold clusters. The scale bar is 10 nm. **(c)** Size distribution of the ligand-protected gold clusters. The mean particle diameter is 1.4 ± 0.6 nm. **(d)** Optical spectra of the Au:DMSA clusters. An absorbance maximum at 390 nm is shown, which is characteristic for Au$_{13}$ cluster.

 **Figure 2. Cluster device and electrical transport behavior of gold cluster.**

**(a)** Schematic illustration of the cluster device structure. **(b)** Experimental I-V and **(c)** dI/dV curves for gold cluster at 1.6 K at different gate voltages. The black, red, blue, pink and green curves correspond to $V_g$=0, 2, 3, 9.5, 11V, respectively. **(d)** Experimental I-V curves measured at different temperatures. The black, red, blue, pink and green curves correspond to T=1.6, 6, 30, 100, 150K, respectively.

**Figure 3. Simulated Structure and Density of States of $Au_{13}(DMSA)_8$ clusters.**

(a) One guess structure of the $Au_{13}(DMSA)_8$ cluster. Orange and yellow atoms represent Au and S atoms, respectively. Grey, red and white dots are for C, H and O atoms. **(b)** The calculated density of states curve and molecular orbital isosurfaces for corresponding peaks. The zero point of the horizontal axis is shifted to the energy level position of the highest occupied molecular orbital (HOMO). The red and blue isosurface is for the HOMO while the pink and light blue isosurfaces are for unoccupied molecular orbitals. The isovalue is 0.04 a.u. (c) Proposed diagram to interpret the electrical measurement.

FIGURES

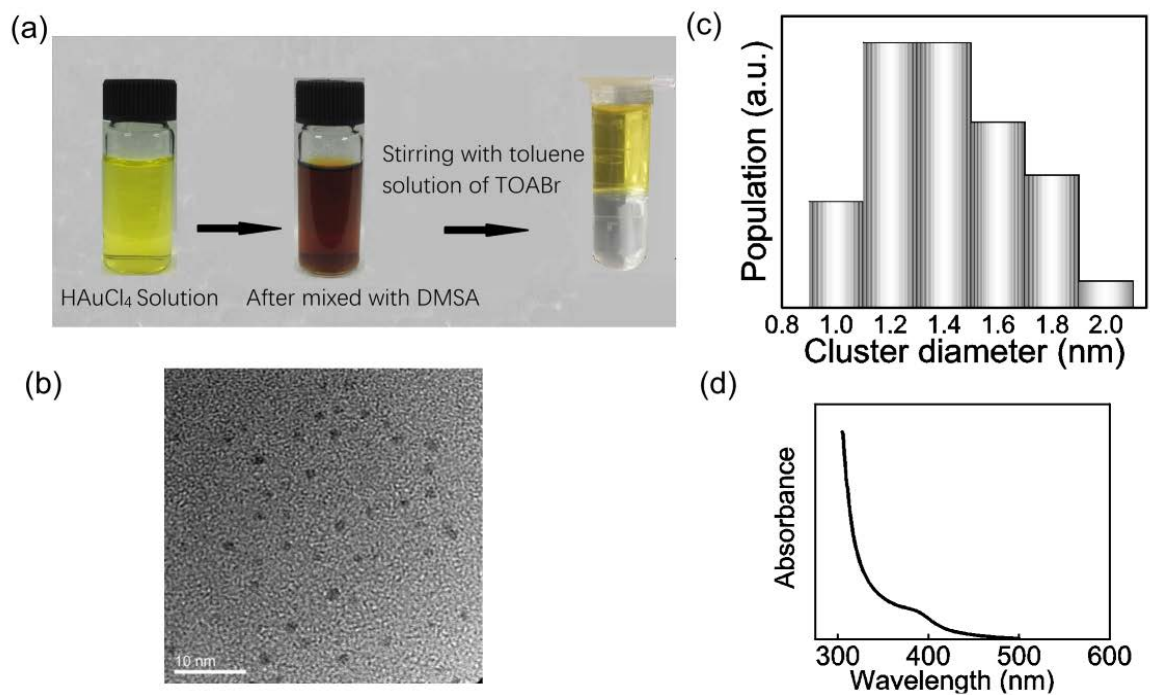

**Figure 1.** Synthesis and Characterization of Au$_{13}$ clusters

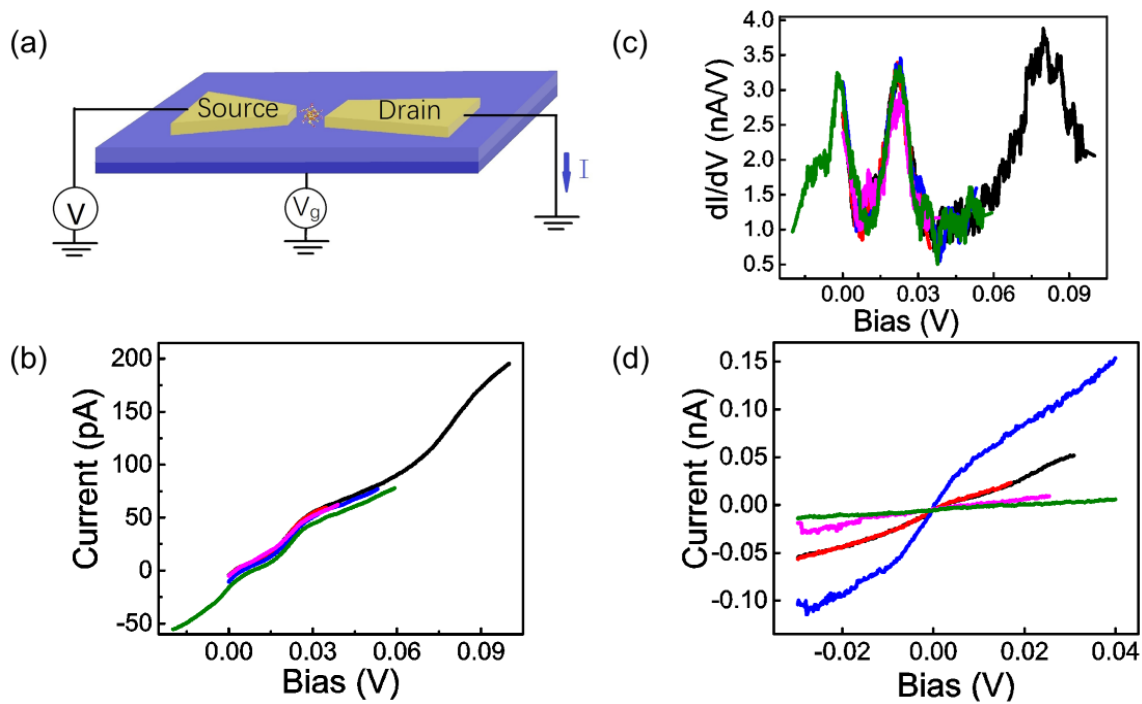

**Figure 2.** Cluster device and electrical transport behavior of gold cluster.

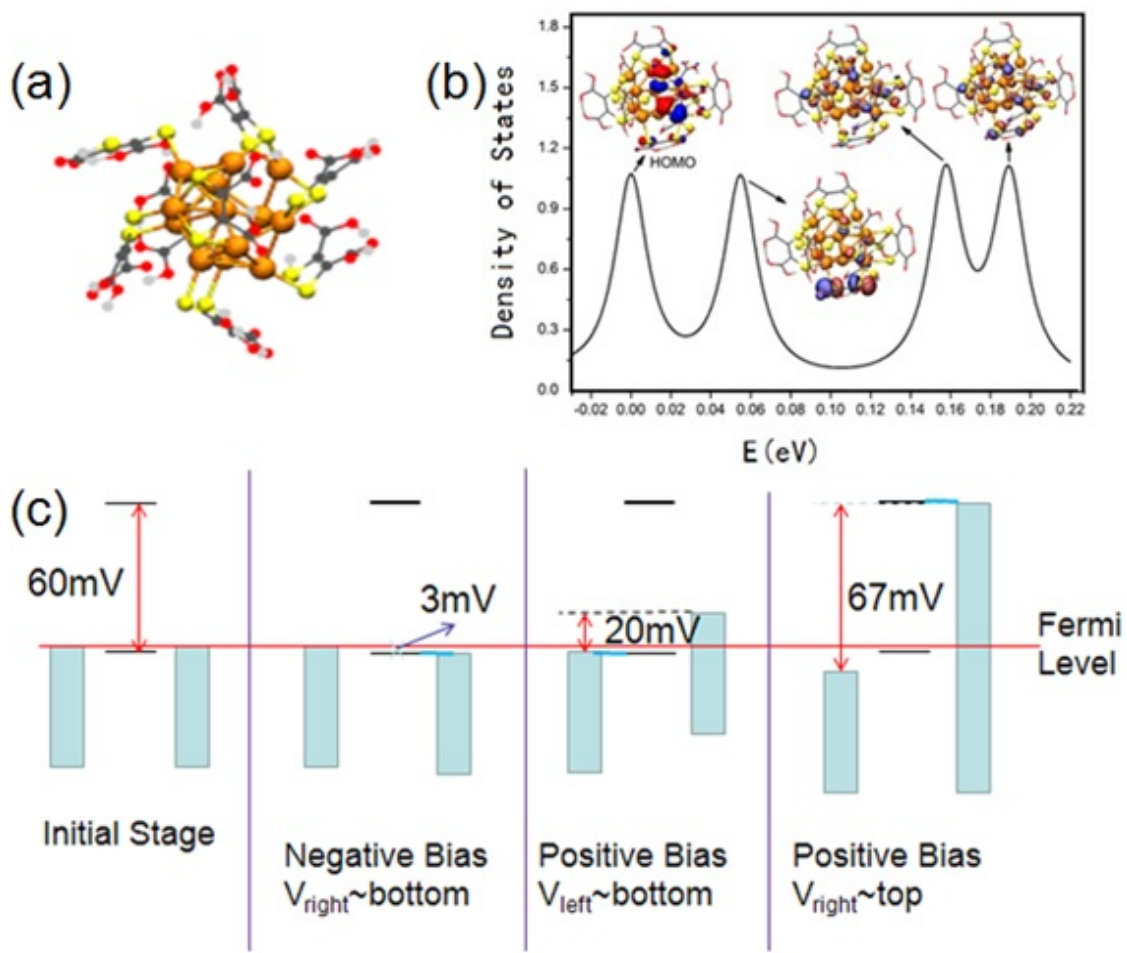

**Figure 3.** Simulated Structure and Density of States of $Au_{13}(DMSA)_8$ clusters.

17